\documentclass[iop,apj]{emulateapj}

\usepackage{times}

\usepackage{amsmath,amssymb,amstext}
\usepackage[breaklinks,colorlinks,citecolor=blue,linkcolor=magenta]{hyperref} 

\usepackage[all]{hypcap} 

\def\D{\mathrm{d}}

\shorttitle{Stellar and Black Hole Mass densities as tracers of co-evolution}

\shortauthors{Schindler et al.}

\begin{document}
 
\title{Stellar and Black Hole Mass densities as empirical tracers of co-evolution \\ show lock-step growth since $z{\sim}3$} 

\author{ Jan-Torge Schindler\altaffilmark{1}, Xiaohui Fan\altaffilmark{1}, Wolfgang J. Duschl \altaffilmark{1,2}}

\altaffiltext{1}{ Steward Observatory, University of Arizona, 933 North Cherry Avenue, Tucson, AZ 85721,USA}
\altaffiltext{2}{Astrophysik Kiel, Institut f\"ur Theoretische Physik und Astrophysik, Christian-Albrechts-Universit\"at zu Kiel, Leibnizstr. 15, 24118 Kiel, Germany}

\begin{abstract}

At redshifts beyond $z{\sim}1$ measuring the black hole galaxy relations proves to be a difficult task. The bright light of the AGN aggravates deconvolution of black hole and galaxy properties. On the other hand high redshift data on these relations is vital to understand in what ways galaxies and black holes co-evolve and in what ways they don't.
In this work we use black hole (BHMDs) and stellar mass densities (SMDs) to constrain the possible co-evolution of black holes with their host galaxies since $z{\sim}5$. The BHMDs are calculated from quasar luminosity functions (QLF) using the Soltan argument, while we use integrals over stellar mass functions (SMFs) or the star formation rate density to obtain values for the stellar mass density.
We find that both quantities grow in lock-step below redshifts of $z\sim3$ with a non-evolving BHMD to SMD ratio.
A fit to the data assuming a power law relation between the BHMD and the SMD and yields exponents around unity ($1.0-1.5$).
Up to $z{\sim}5$ the BHMD to SMD ratio doesn't show a strong evolution given the larger uncertainty in the completeness of high-redshift datasets. 
Our results, always applying the same analysis technique, seem to be consistent across all adopted data sets.

\end{abstract}


\section{Introduction}

Tight correlations between the masses of super massive black holes (SMBH; $M_\bullet > 10^6\,M_\odot$) and properties of the stellar component of their host galaxies are well established. These connect stellar bulge properties, such as the velocity dispersion $\sigma_{\rm{sph}}$ \citep[e.g.,][]{Ferrarese2000, Gebhardt2000,Tremaine2002, Gueltekin2009, Graham2011}, the luminosity $L_{\rm{sph}}$ \citep[e.g.,][]{Kormendy1995, McLure2001, Gueltekin2009, Schulze2011, Kormendy2013} and the stellar bulge mass $M_{\star,\rm{bulge}}$ \citep[e.g.,][]{Magorrian1998, McLure2002, Marconi2003, Haering2004, Beifiori2012} with the SMBH mass $M_\bullet$.

The existence of these relations support the idea of a coordinated co-evolution of SMBHs and their host galaxies by a common physical mechanism. 

One mechanism proposed to explain this relates black holes (BH) and stellar mass growth via feedback processes from active galactic nuclei (AGN) through momentum or energy-driven winds \citep{Silk1998, Fabian1999, King2003,Wyithe2003, Murray2005, DiMatteo2005, Croton2006, Bower2006, Hopkins2006, Somerville2008}. 

On the other hand, \citet{Peng2007} and \citet{Jahnke2011} demonstrated that merger averaging of galaxy stellar and SMBH mass creates the correlation between both quantities via a statistical convergence process rather than a physical process. 
Assuming that BH masses and stellar masses have initially uncorrelated distributions, hierarchical merging can create the scaling relations we see today without invoking any kind of physical mechanism connecting both quantities.  
In this framework a need for a physical connection between the bulge and the SMBH via AGN feedback or other mechanisms does not seem to be necessary.

Although these correlations have been known for more than a decade, the dominant process that engineered them is still under debate \citep[see][for a discussion]{Kormendy2013}.

The problem is compounded at high redshifts where BH masses cannot be measured directly, but must be inferred from emission line widths. Furthermore, the strong emission of the nucleus at large distances severely limits the ability to measure host galaxy properties. Measuring the bulge stellar mass in those systems requires color information, which is mingled with the spectrum of the bright nucleus. Velocity dispersion measurements suffer from the presence of AGN emission lines. Studies of the $M_\bullet-\sigma$ relation have been carried out up to redshifts of $z\sim0.6$ \citep[e.g.][]{Woo2008, Canalizo2012, Harris2012, Hiner2012}, while studies of the $M_\bullet-M_{\star,\rm{bulge}}$ relation reached redshifts up to $z\sim6$ \citep[e.g.][]{Walter2004, Wang2010, Targett2012}.
These studies at high redshift have very limited sample sizes; many also used potentially biased tracers (e.g., the most luminous quasars) for the BH populations. They might not represent the full galaxy and BH populations at high redshift. 

To overcome the systematics that are associated with sample selection and the difficulties with high redshift data, we turn towards a more statistical approach to analyze the prospect of BH and galaxy co-evolution.
In this study we focus on the global properties of the whole galaxy and BH population by calculating the evolution of the total stellar mass and BH mass density (SMD and BHMD, respectively) over redshifts $z=0-5$. 

This method has been applied before by \citet{Zhang2012} at $z\leq1.2$. At these redshifts it is possible to estimate not only the total stellar mass density but the bulge stellar mass density from bulge to total mass ratios and morphology distributions. 

However, at redshifts beyond $z\approx2$ galaxies appear to be significantly different \citep{vanDokkum2008} and bulge to total mass ratios as well as morphology distributions are unconstrained by observations. Therefore we calculate the total stellar mass density (SMD) rather than the bulge mass density.

\citet{Laesker2014b} investigated the $M_\bullet-L_{\rm{sph}}$ relation using deep K-band photometry finding a linearly proportional relation between SMBH masses and total galaxy luminosity ($M_\bullet-L_{\rm{tot}}$) with comparable scatter to the $M_\bullet-L_{\rm{sph}}$ relation. The authors suggest that the total instead of the bulge luminosity (mass) might be the driving factor in the co-evolution of SMBHs and their hosts.

More recently \citet{Reines2015} investigated the local $M_\bullet-M_{\star,\rm{host}}$ relation in a sample of broad-line AGN in the local universe, complementary to \citet{Laesker2014b}, finding a clear correlation between BH mass and total stellar mass.
The normalization of this relation is, however, by an order of magnitude lower.

We calculate the SMD in two distinct ways, first using stellar mass functions from various studies, secondly adopting the star formation rate density of \citet{Madau2014} and integrating it across cosmic time.

The BHMD is estimated using quasar luminosity functions (QLFs) as tracers of the accretion history of bright quasar phases. \citet{Soltan1982} first put forward the argument that the BHMD from bright quasars phases can be inferred from the energy density of photons via a mass-to-light energy conversion efficiency \citep[see also][]{Chokshi1992}.

We devote Section\,\ref{sec_SMD} to derive the SMDs including uncertainties and compare our results with estimates from other authors. In Section\,\ref{sec_BHMD} we continue with the calculation of the BHMD in a similar fashion. We then proceed with the analysis of the combined data in Section\,\ref{sec_relation} and present the relation between both quantities derived in the previous chapters. Section\,\ref{sec_discussion} discusses implications of the presented cosmic evolution of SMD and BHMD and compares our results to those of the EAGLE and Illustris simulations. We summarize our findings in Section\,\ref{sec_conclusion}.

For the calculations we adopt the cosmology that the SMFs and QLFs were determined in, which is mostly a $\Lambda$-CDM cosmology parameterized by $h=0.7$, $\Omega_\Lambda=0.7$ and $\Omega_{\rm{M}}=0.3$. Only in the case of the QLF of \citet{Palanque2013} the cosmological parameters are slightly different, $h=0.71$, $\Omega_\Lambda=0.734$ and $\Omega_{\rm{M}}=0.267$. This resulting difference in the BHMD of $5.8\%$ at z=3 is much smaller than the $1\sigma$ uncertainties at this redshift and thus negligible.

\section{The cosmic stellar mass density (SMD)} \label{sec_SMD}

\subsection{Calculating the SMD Using Stellar Mass Functions}

A stellar mass function $\Phi_{\rm{SMF}}(z)$ (SMF) describes the number density of galaxies $n_{\rm{gal}}$ per mass interval ($M_{\rm{gal}},M_{\rm{gal}}+\D M_{\rm{gal}})$ in a cosmic volume at a given redshift $z$. 

SMFs are built by estimating the individual masses of galaxies in large surveys via population synthesis modeling. Assumptions about the galaxy (e.g. initial mass function (IMF), star formation history, stellar metallicity distribution, age, etc.) and its stellar mass are used to build model SEDs that are fitted to the actual galaxy SED. The best fit yields a stellar mass estimate for each galaxy, which are then compiled to build a SMF for the full galaxy sample. Note that the choice of IMF has a strong impact on the SMF.

Stellar mass functions are most commonly described by single or double Schechter functions \citep{Schechter1976,Baldry2008}. They consist of a single/double power law in mass that turns into a falling exponential above a characteristic mass scale $M^*$. 

Most often the stellar mass distributions are divided into redshift bins for which the best fit parameters are estimated assuming one of the above functional forms.

Once the SMF is defined, the SMD $\rho_{\rm{gal,\star}}(z)$ between the stellar mass integration limits, $M_{\rm{gal,\star}}$ and $M_{\rm{gal,\star}} + \Delta$, can then be computed by a simple integration,

\begin{equation}
 \rho_{\rm{gal,\star}}(z) = \int_{M_{\rm{gal,\star}}}^{M_{\rm{gal,\star}} + \Delta} M \Phi_{\rm{SMF}}(M,z) \D M \label{eq_calcSMD} \ .
\end{equation}

\subsection{Stellar Mass Functions}

We use four recent measurements of the SMF to calculate the SMD in all available redshift bins. A summary of the most important characteristics is provided in Table\,\ref{tab_smfsummary}. For our purpose we rescaled all calculated SMD values to a Salpeter IMF.

For near up to intermediate redshifts ($z<4$) we used the SMFs of \citet{Bielby2012, Ilbert2013} and \citet{Muzzin2013}. 
The two latter works use data from the UltraVista/COSMOS field to determine quiescent and star-forming SMFs in the redshift range of $0.2 < z < 4.0$. The UltraVista/COSMOS is the largest galaxy dataset to date and therefore provides the most complete estimates of the SMF. Although both works  base their SMFs on a similar data set the authors employ different analysis methods. We deliberately include both SMFs in our analysis to use the robustness of the underlying sample to derive the SMDs and furthermore display the effects that different analysis methods have on the SMDs. We use the best fit parameters of  \citet[their Table\,2]{Ilbert2013} and \citet[their Table\,7]{Muzzin2013}.

The SMF of \citet[their Table\,7]{Bielby2012} is based on data from the four Canada-France-Hawaii Telescope Legacy Survey (CFHTLS) deep fields augmented by new near-infrared data from the WIRCam Deep Survey. In comparison to most other SMF estimates the data from this sample is not encompassed in the UltraVista/COSMOS fields and therefore independent of the SMF of \citet{Ilbert2013} and \citet{Muzzin2013}. The four different fields help to reduce uncertainties due to cosmic line of sight variations. The SMF is determined over a redshift range of $0.2 < z < 2.0$.

At higher redshifts ($z>4$) we adopt the SMF of \citet[their Table\,2]{Caputi2011} which is determined at redshifts $3 < z < 5$ using data from the UKIRT Infrared Deep Survey (UKIDSS) Ultra Deep Survey (UDS). This study's IRAC galaxy selection avoids the potential bias of UV selection against older or dustier galaxies. However, the depth of the IRAC data limits the direct detections to a higher stellar mass completeness limit.

\subsection{Integration Limits}
For the high mass integration limit in Eq.\,\ref{eq_calcSMD} we adopt a value of $10^{13}\,M_\odot$. Due to the exponential decline of the SMF at high masses the SMDs are complete at the $0.001\%$ level if this value is adopted. It is also a common value in the literature  \citep[e.g.][]{Caputi2011,Ilbert2013, Muzzin2013} which again ensures comparability of our calculated SMDs.

The low mass limit of the SMFs faces completeness issues because of the survey's limited sensitivity. Below a certain flux limit galaxies simply cannot be detected anymore. Mass completeness is not as well defined as luminosity completeness due to the broad range of mass to light ratio values that galaxies can exhibit. We decided to use a common generic lower mass limit of $10^8\,M_\odot$ to make our results comparable to other studies \citep[e.g.][]{Caputi2011,Ilbert2013, Muzzin2013}. Unfortunately this mass limit does not represent the sensitivities of the different SMFs and thus introduces systematics by extrapolating below the mass completeness limit.
However, since the low mass slopes of the SMFs are flatter than of galaxy luminosity functions, the impact of the lower mass limit on the SMD is not as big as it would be for the star formation rate density.
To test the robustness we varied the lower integration boundary between$10^7$ to $10^9\,M_\odot$ and found that the SMDs of two of the SMFs \citep{Ilbert2013, Muzzin2013} are quite robust against this variation, while the other \citep{Bielby2012, Caputi2011} do show a non negligible dependence on the lower mass integration limit. Hence adopting the value of $10^8\,M_\odot$ as our lower mass integration limit, could bias our results of the SMD. 

\subsection{Error estimation}

We estimate our uncertainties on the SMDs by using a Monte Carlo method. 
For each parameter of the SMF we build a Gaussian distribution with the reported mean and $1\sigma$ uncertainty. Whenever asymmetric uncertainties were quoted, we use the average of the upper and lower uncertainty as our symmetric $1\sigma$ uncertainty for the Gaussian, because the multi dimensional distribution functions of all parameters were not available.

For each realization of the SMF all parameters are randomly drawn from these distributions. Using Eq.\ref{eq_calcSMD} we calculate the SMD from this randomly drawn SMF.
After calculating ten thousand realizations of the SMD per SMF per redshift bin, we build cumulative distributions of the stellar mass densities and fit them with a cumulative Gaussian distribution to derive the mean and the $1\sigma$ uncertainties.

\subsection{Stellar Mass Densities from Stellar Mass Functions}

\begin{figure}[ht]
\includegraphics[width=0.45\textwidth]{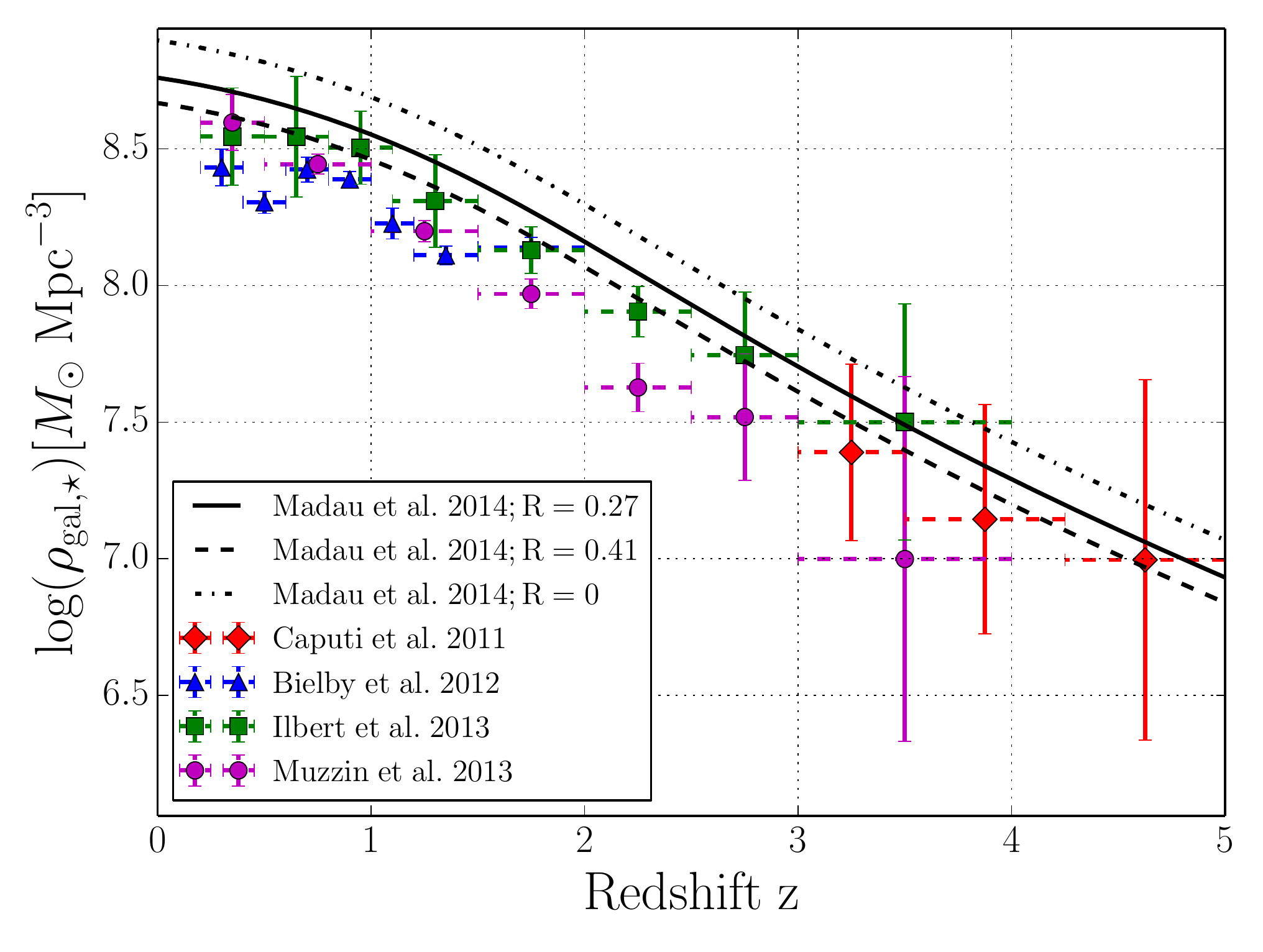}
\caption{The colored dots and error bars along the ordinate display the Monte Carlo mean SMD and the corresponding $1\sigma$ uncertainties. The error bars along the abscissa show the redshift bins. The black lines show the SMD based on the cosmic SFRD of \citet{Madau2014} with various values for the return fraction $R$.}
\label{figure_smd}
\end{figure}

We present all calculated SMDs in Figure\,\ref{figure_smd}. Below a $z\approx1$ the cosmic SMD values seem to level off. Above this redshift all SMDs start to decline towards higher redshifts while the uncertainties in the SMD increase strongly at redshifts beyond $z>3$ in all adopted SMFs. The error bars along the redshift axis represent the redshift bins in which the original SMFs were determined.

In Table\,\ref{tab_SMDresults} we show our results of the SMD and compare them with the results quoted in \citet{Madau2014}. Our main results have uncertainties obtained in the Monte Carlo process outlined above. They are shown in the first column of the table. The second column shows the SMD based on the best fit mean parameters of the SMF. The third column contains the SMD values from \citet{Madau2014} to compare with.

Our Monte Carlo results match the literature values well at at all but the highest redshift bins. The values for the SMD of \citet{Bielby2012, Ilbert2013, Muzzin2013} were rescaled to a Salpeter IMF and we believe that the differences to the values of \citet{Madau2014}, which are on the order of $0.2-0.3\,\rm{M}_\odot\, \rm{Mpc}^{-3}$, are due to this process. 
The uncertainties on our Monte Carlo SMD values are higher since the covariances of the best fit SMF parameters were not included in the publication. Therefore our Monte Carlo approach delivers a more conservative estimate of the uncertainties.

\begin{table}[ht]
 \centering
 \caption{The Calculated SMDs Compared to Values from the Literature.}
 \label{tab_SMDresults}

\begin{tabular}{cccc}
\tableline 
\tableline \\
 Redshift &\multicolumn{3}{c}{$\log\rho_{\rm{gal,\star}} [\rm{M}_\odot\, \rm{Mpc}^{-3}]$} \\
       & \footnotesize  Based on  &\footnotesize  Based on mean &\footnotesize  Comparison \\
       & \footnotesize   Monte Carlo fit  &\footnotesize SMF parameters &\footnotesize \citet{Madau2014} \\
 \tableline
& \multicolumn{2}{c}{of SMF \citep{Caputi2011} } & \\
 \tableline \\
 $3.00 - 3.50$ & $7.33\pm0.35$ & $7.32$ & $7.32^{+0.04}_{-0.02}$   \\ 
 $3.50 - 4.25$ & $7.08\pm0.44$ & $7.05$ & $7.05^{+0.11}_{-0.10}$   \\ 
$4.25 - 5.00$ & $6.96\pm0.68$ & $6.38$ & $6.37^{+0.14}_{-0.54} $   \\ 
 \tableline
& \multicolumn{2}{c}{of SMF \citep{Bielby2012} } & \\
\tableline \\
 $0.20 - 0.40$ & $8.43\pm0.07$& $8.42$  & $8.46^{+0.09}_{-0.12}$ \\
 $0.40 - 0.60$ & $8.30\pm0.04$& $8.30$  & $8.33^{+0.03}_{-0.03}$ \\
 $0.60 - 0.80$ & $8.43\pm0.05$& $8.41$  & $8.45^{+0.08}_{-0.10}$ \\
 $0.80 - 1.00$ & $8.39\pm0.03$& $8.39$  & $8.42^{+0.05}_{-0.06}$ \\
 $1.00 - 1.20$ & $8.23\pm0.06$& $8.22$  & $8.25^{+0.04}_{-0.04}$ \\
 $1.20 - 1.50$ & $8.11\pm0.03$& $8.11$  & $8.14^{+0.06}_{-0.06}$ \\
 $1.50 - 2.00$ & $8.14\pm0.04$& $8.14$  & $8.16^{+0.32}_{-0.03}$ \\
 \tableline
& \multicolumn{2}{c}{of SMF \citep{Ilbert2013} } & \\
\tableline \\
 $0.20 - 0.50$ & $8.53\pm0.18$ & $8.52$ & $8.55^{+0.08}_{-0.09}$ \\
 $0.50 - 0.80$ & $8.56\pm0.23$ & $8.44$ & $8.47^{+0.07}_{-0.08}$ \\
 $0.80 - 1.10$ & $8.51\pm0.14$ & $8.47$ & $8.50^{+0.08}_{-0.08}$ \\
 $1.10 - 1.50$ & $8.31\pm0.17$ & $8.30$ & $8.34^{+0.10}_{-0.07}$ \\
 $1.50 - 2.00$ & $8.13\pm0.09$ & $8.12$ & $8.11^{+0.05}_{-0.06}$ \\
 $2.00 - 2.50$ & $7.90\pm0.09$ & $7.90$ & $7.87^{+0.08}_{-0.08}$ \\
 $2.50 - 3.00$ & $7.73\pm0.24$ & $7.70$ & $7.64^{+0.15}_{-0.14}$ \\
 $3.00 - 4.00$ & $7.45\pm0.45$ & $7.33$ & $7.24^{+0.18}_{-0.20}$ \\
 \tableline
& \multicolumn{2}{c}{of SMF \citep{Muzzin2013} } & \\
\tableline \\
 $0.20 - 0.50$ & $8.59\pm0.10$ & $8.59$ & $8.61^{+0.06}_{-0.06}$ \\
 $0.50 - 1.00$ & $8.44\pm0.04$ & $8.44$ & $8.46^{+0.03}_{-0.03}$ \\
 $1.00 - 1.50$ & $8.20\pm0.04$ & $8.20$ & $8.22^{+0.03}_{-0.03}$ \\
 $1.50 - 2.00$ & $7.97\pm0.05$ & $7.97$ & $7.99^{+0.05}_{-0.03}$ \\
 $2.00 - 2.50$ & $7.62\pm0.09$ & $7.62$ & $7.63^{+0.11}_{-0.04}$ \\
 $2.50 - 3.00$ & $7.52\pm0.23$ & $7.50$ & $7.52^{+0.13}_{-0.09}$ \\
 $3.00 - 4.00$ & $7.05\pm0.72$ & $6.83$ & $6.84^{+0.43}_{-0.19}$ \\
\tableline \\
\multicolumn{4}{c}{ 
\begin{minipage}{1.\columnwidth}
\footnotesize \textbf{Notes.} All SMD values are rescaled to a Salpeter IMF, where it was necessary. 
\end{minipage}}
\end{tabular}
\end{table}

\subsection{Stellar Mass Density derived From the Instantaneous Star Formation Rate Density}

We also infer the SMD from the best fit instantaneous star formation rate density (SFRD) $\Psi(z)$ given by Eq.\,15 in \citet{Madau2014}. Their best fit SFRD considers measured star formation rates (SFRs) in the rest-frame far UV, mid IR and far IR (see their Table\,1). We integrate $\Psi(z)$ from $z=10$ to $z=0$ using their Eq.\,2 with the return fraction $R$. The return fraction is a measure of how much stellar mass is returned to the interstellar medium during the stellar population life cycle. It is one major uncertainty from converting the SFRD to SMD. We show the range of values for the SMD by using three different values for the return fraction, $R=0.27$ for a Salpeter IMF, $R=0.41$ for a Chabrier IMF and a value of $R=0$ assuming no material is returned to the interstellar medium. The resulting three curves are shown as the solid, the dashed and the dashed-dotted black lines in Fig.\,\ref{figure_smd}. They illustrate the uncertainty associated with the SMD calculated from the SFRD. For all further purposes we use the SMD with the return fraction of $R=0.27$, which reflects a Salpeter IMF. \citet{Madau2014} state their SMD calculated from their SFRD is somewhat higher ($\sim 0.5\,\rm{dex}$) than the SMD data from SMFs at all redshifts. For a discussion about this discrepancy we refer to the original study. 
The SMD calculated from the SFRD provides a SMD measurement independent to the SMFs used above.

\section{The Cosmic Black Hole Mass density Inferred from QSOs} \label{sec_BHMD}

\subsection{Calculating the Cosmic BHMD Using the Soltan Argument}

Individual black holes can grow via mass accretion or in mergers. In the latter process some black hole mass can be lost due to the emission of gravitational waves. 
The BHMD, assuming that mass loss during mergers is negligible, can only grow by mass accretion. Merger processes do not change the total black hole mass in a large enough volume of space. 

We can write the total accreted black hole mass density $\rho_{\bullet,\rm{acc}}(z)$ between redshifts $z_0$ and $z$ as a function of the total black hole mass growth rate $\dot{M}_{\bullet,\rm{acc}}$ in a volume of space $V$,
\begin{equation}
 \rho_{\bullet,\rm{acc}}(z) \equiv \int_{z_0}^z \frac{\dot{M}_{\rm{\bullet, acc}}(z')}{V} \left(\frac{\D t(z')}{\D z'}\right) \D z' \label{eq_BHMDtoBHMgrowth} \ .
\end{equation}
In order to substitute the black hole mass growth rate with an observational quantity we adopt the Soltan argument \citep{Soltan1982} and assume that black holes grow only by mass accretion during bright AGN phases. The total bolometric luminosity of all accreting black holes in a volume of space $L_{\rm{bol},\rm{tot}}(z)$ can then be directly related to their overall accretion rate $\dot{M}_{\rm{acc}}(z)$, introducing an average efficiency factor of energy conversion $\epsilon$. This factor also relates the overall accretion rate with the total black hole mass growth rate $\dot{M}_{\bullet,\rm{acc}}$.
\begin{equation}
 L_{\rm{bol},\rm{tot}}(z) = \epsilon\ \dot{M}_{\rm{acc}}(z)\ c^2 = \frac{\epsilon\ c^2}{1-\epsilon} \dot{M}_{\bullet,\rm{acc}} \label{eq_LboltoBHMgrowth}
\end{equation}
Inserting Eq.\ref{eq_LboltoBHMgrowth} into Eq.\ref{eq_BHMDtoBHMgrowth} and substituting the bolometric luminosity per cosmic volume with an integral over the quasar luminosity function $\Phi_{\rm{QLF}}$ we can now calculate the BHMD,
\begin{align}
\nonumber \rho_{\bullet,\rm{acc}}(z)=&  \frac{1-\epsilon}{\epsilon\ c^2} \int_{z_0}^z \left(\frac{\D t(z')}{\D z'}\right)\\ &  \int_{L}^{L + \Delta} L_{\rm{bol}}(L) \Phi_{\rm{QLF}}(L, z')  \D L\ \D z' \label{eq_calcBHMD} \ .
\end{align}
The lower and upper luminosity integration limits are denoted by $L$ and $L + \Delta$.
This expression and the derivation have been applied many times in previous works \citep[e.g][]{Yu2002,LaFranca2005,Hopkins2007,Zhang2012}. 
We do not include the BH seed masses in our definition of the BHMD. Their contribution to the BHMD at redshifts around and below $z=5$, is negligible since the accreted BH mass in bright quasar phases will be larger by orders of magnitude compared to the BH seed mass.

\subsection{Efficiency Factor of Energy Conversion}

For a rotating black hole the efficiency factor of energy conversion varies between $\epsilon \sim 0.06$ from a non-rotating black hole to $\epsilon \sim 0.31$ for a maximally rotating Kerr black hole \citep{Thorne1974}. Only theoretical arguments currently constrain the distribution function of black hole spins and therefore the distribution of efficiency factors. Above we assume an average efficiency factor for the whole black hole population allowing us to write $\epsilon/(1-\epsilon)$ in front of the integral in Eq.\ref{eq_calcBHMD} as a scaling factor. Whenever we calculate values for the BHMD we adopt a value of $\epsilon/(1-\epsilon) = 0.1$ to be comparable with previous works \citep{Yu2002,Ueda2003,Marconi2004,Hopkins2007,Zhang2012}.

\subsection{Quasar Luminosity Functions}
The calculated BHMD strongly depends on the adopted quasar luminosity functions (QLFs) and their redshift evolution. Most QLF are parameterized by a double power law,
\begin{equation}
 \Phi_{\rm{QLF}} = \frac{\D n_{\rm{QSO}}}{\D \log L} =  \frac{\Phi^*}{(L/L^*)^\alpha+(L/L^*)^\beta} \ ,
\end{equation}
where $\Phi^*$ is the normalization constant and $L^*$ is the characteristic luminosity where the power law index changes from $\alpha$ to $\beta$. 
The QLF yields the number density of quasars $n_{\rm{QSO}}$ of a certain luminosity range if integrated over it. This distribution function, an important statistic for active galactic nuclei, is determined from wide surveys with large numbers of objects at various redshifts. 

The QLF is often fitted with a continuous redshift evolution in all or a subset of its parameters $\alpha, \beta, L^*$ and $\Phi^*$.
If only the characteristic luminosity $L^*$ is varied as a function of redshift, the evolution is termed the pure luminosity evolution (PLE) model. If characteristic luminosity and normalization constant $\Phi^*$ are parameterized in redshift but independent from each other, one speaks of a luminosity evolution and density evolution model (LEDE). If the density evolution depends on the characteristic luminosity, the model is called luminosity dependent density evolution (LDDE). Some authors also vary both slopes or introduce a pivot redshift above and below which the slopes have different values. 
We calculate six different BHMD evolutions to compare with the SMDs based on the following QLFs. A summary of their most important characteristics is provided in Table\,\ref{tab_qlfsummary}.

\textbf{a)} \citet{Hopkins2007} combined a large set of QLF measurements, from the rest-frame optical, soft and hard X-ray, and near- and mid-IR bands to determine a bolometric QLF. Their bolometric QLF spans a redshift range of $ z= 0-6$. The QLF is best fit by a PLE model with an additional redshift dependence of the faint and bright end slopes. The best fit values are given in their Table\,3 of \citet[``full'' model]{Hopkins2007}. The large set of QLFs folded into the determination of this bolometric QLF overcomes the bias against obscured objects in the optical wavelength range that optical single waveband QLFs suffer from.
We further adopted three optical QLFs in the g- and i-band.

\textbf{b)} \citet{Croom2009} compiled a QLF from 10637 QSOs of the 2dF-SDSS LRG and QSO survey in the redshift range $0.4 < z < 2.6$. Their g-band magnitudes reach a flux limit of 21.85 and an absolute continuum magnitude of $M_{\rm{g}}(z=2) < -21.5$. 

\textbf{c)} The QLF of \citet{Ross2013}, $0.3 < z < 3.5$, measured using 22301 quasars in the i-band of the SDSS:BOSS, serves as a second optical QLF with a large number of objects over a wide field. 

\textbf{d)} To extend the g-band data we also included the QLF of \citet{Palanque2013} in our analysis. Their QLF, measured in the rest frame g-band, complementing Croom et al.'s QLF with a wider redshift range, $0.68 < z < 4.0$, but with the downfall of a smaller sample size and narrower area on the sky.

We use the best fit modified LEDE model of \citet[their Table\,4]{Croom2009}, the PLE model with the pivot redshift of \citet[their Table\,7]{Palanque2013} and the first PLE model (redshift range $0.3 < z < 2.2$) in addition to the second LEDE model (DR9) of \citet[their Table\,8]{Ross2013}.

\textbf{e)} We further used the combined hard X-ray QLFs of \citet{LaFranca2005} and \citet{Fiore2012}. 
In the low redshift regime ($z\leq 3.0$) we use the QLF of \citet{LaFranca2005} which was fit using a total of 508 AGN from the HELLAS2XMM sample and other published catalogs (see their Table\,1). 
\citet{Fiore2012} set out to extend the results of \citet{LaFranca2005} and placed constraints on the QLF at redshifts $3 < z < 7$ using new 4Msec Chandra observations in the Chandra Deep Field South (CDFS). Their QLF fit follows the usual double power law with a combination of luminosity and density evolution model (termed LADE), fully equivalent to the introduced LDDE model above. We use the best fit model (4) of \citet[their Table\,5]{Fiore2012} joined to the best fit model (LDDE) of \citet[][their Table 2]{LaFranca2005} at a redshift of $z=3$ to determine the BHMD. 

\textbf{f)} At last we adopt another estimate of the hard X-ray QLF by \citet{Ueda2014}.
The authors of this study have compiled a very large sample of 4039 AGN detected between redshifts $z=0-5$ from a multitude of surveys (see their Table\,1). They perform a maximum likelihood method to estimate a bolometric QLF fit, parameterized using a luminosity-dependent density evolution. We use the parameters given in their Table\,4 (``Bolometric'') to calculate the BHMD for this QLF.

\subsection{K-correction and Bolometric Correction}
QLFs are a direct product of quasar fluxes observed in a certain waveband. Therefore their luminosities are wavelength dependent. In order to calculate the BHMD we need to convert the rest frame luminosity (or magnitude) to a bolometric luminosity (see Eq.\ref{eq_calcBHMD}). 
However, many authors follow \citet{Richards2006} and apply a continuum K-correction to their QLFs relative to a redshift of $z_0=2$. 
Hence, before we can convert the waveband dependent luminosity (or magnitudes) to the bolometric luminosity we have to apply a K-correction relative to a redshift of $z=0$ to all optical QLFs. The K-correction is waveband independent for power-law spectral energy distributions (SED). The correction for magnitudes takes the following form \citep[see][]{Richards2006},
\begin{equation}
 M(z=0) = M(z=z_0) + 2.5 (1+\alpha_{\nu}) \log (1+z_0) \ .
\end{equation}
The power law slope of the SED is denoted by $\alpha_{\nu}$.

After applying the K-correction we can now use the luminosity dependent bolometric correction of \citet[Eqs.\,2 and 3]{Hopkins2007} for the B-band and the Hard X-ray to convert the optical and X-ray QLFs to bolometric luminosities.

Some of our adopted QLFs are calculated in the g- or the i-band. We use the conversions given in \citet{Ross2013} and \citet{Jester2005} to formulate them in B-band luminosities to be able to apply the bolometric corrections correctly.

\subsection{Integration limits}

In order to numerically solve Eq.\,\ref{eq_calcBHMD} one needs to set the integration limits for the luminosity as well as for the redshift evolution. 

To ensure that all BHMDs are comparable with another and correspond well to galaxy masses in the range of $10^8\,M_\odot$ to $10^{13}\,M_\odot$, we adopted uniform lower and upper integration limits for the luminosity. In bolometric luminosities these limits are $10^8 L_\odot$ and $10^{18} L_\odot$. They lie beyond some of the magnitude/luminosity completeness-limits of the QLFs and thus we necessarily extrapolate the QLFs in these cases. The individual integration limits for all QLFs are given in Table\,\ref{tab_qlfintlim}.

All QLFs are necessarily extrapolated to higher redshifts where no data constrains them. However, the mass in the extrapolated regime is small compared to the BH mass growth below $z\sim5$. 
Assuming two different redshifts $z_0=10, 20 $ for the start of the redshift evolution we examined the changes in the BHMD for all QLFs at a redshift of $z=5$. We found no changes at a $0.001\%$ level between $z_0=10$ and $z_0=20$. Hence, we decide to start our integration at redshift $z_0 = 10$.

\begin{table}
\scriptsize
\centering
\caption{Integration limits for the QLF in the BHMD calculation (Eq.\,\ref{eq_calcBHMD})}
\label{tab_qlfintlim}
\begin{tabular}{cc}
\tableline
\tableline
 QLF Reference & QLF Integration Limits \\
\tableline
\citet{Hopkins2007} & $L_{\rm{bol}} = 10^8$ to $10^{18}\,L_\odot$ \\
\citet{Croom2009} & $M_{\rm{g}}(z=2) =-39.00$ to $-12.01$ \\
\citet{Palanque2013} & $M_{\rm{g}}(z=2) =-38.77$ to $-11.78$ \\
\citet{Ross2013} & $M_{\rm{i}}(z=2) =-38.84$ to $-11.85$ \\
\citet{LaFranca2005} & $\log L_{\rm{hX}}[\rm{erg}\,\rm{s}^{-1}] = 40.60$ to $48.31$\\
\citet{Fiore2012} & $\log L_{\rm{hX}}[\rm{erg}\,\rm{s}^{-1}] = 40.60$ to $48.31$ \\
\citet{Ueda2014} & $\log L_{\rm{bol}}[\rm{erg}\,\rm{s}^{-1}] = 41.58$ to $51.58$ \\
\tableline
\end{tabular}
\end{table}

\subsection{Correcting for Observable Quasar Population}

In order to convert an observed QLF to a bolometric QLF one also has to take in account extinction effects. 
In their analysis \citet{Hopkins2007} determined the probability of observing quasars with an intrinsic luminosity $L_{\rm{bol}}$ at some observed luminosity $L$. They expressed this probability in an ``observable fraction'' $f(L)$ \citep[Eq.\,4]{Hopkins2007} with a simple power law for conversion between the intrinsic population and the observed one,
\begin{equation}
 f(L) = \frac{\Phi(L)}{\Phi[L_{\rm{bol}}(L)]} = f_{46} \left( \frac{L_{\rm{bol}}}{10^{46}\rm{erg}\rm{s}^{-1}} \right)^\beta \ .
\end{equation}
We adopt this rough approximation for our work to properly convert the observed QLFs to bolometric QLFs taking in account the obscured quasar population with the factor $f(L)$. For our hard X-ray QLFs we use their corresponding values ($f_{46}=1.243$, $\beta=0.066$)  while we use their values for the B-band ($f_{46}=0.26$, $\beta=0.082$) for all optical QLFs.

\subsection{Error estimation}

In order to estimate the uncertainties on the BHMD we apply the same Monte Carlo approach that we used for the SMDs. 
We not only draw random values for all QLF parameters but also for the bolometric correction using the quoted uncertainties in \citet{Hopkins2007}. We do not account for uncertainties in the correction for the observable fraction. From the ten thousand realizations we calculate the median value and the 68 percentile regions.

\subsection{The calculated BHMD}

\begin{figure}[h]
\includegraphics[width=0.5\textwidth]{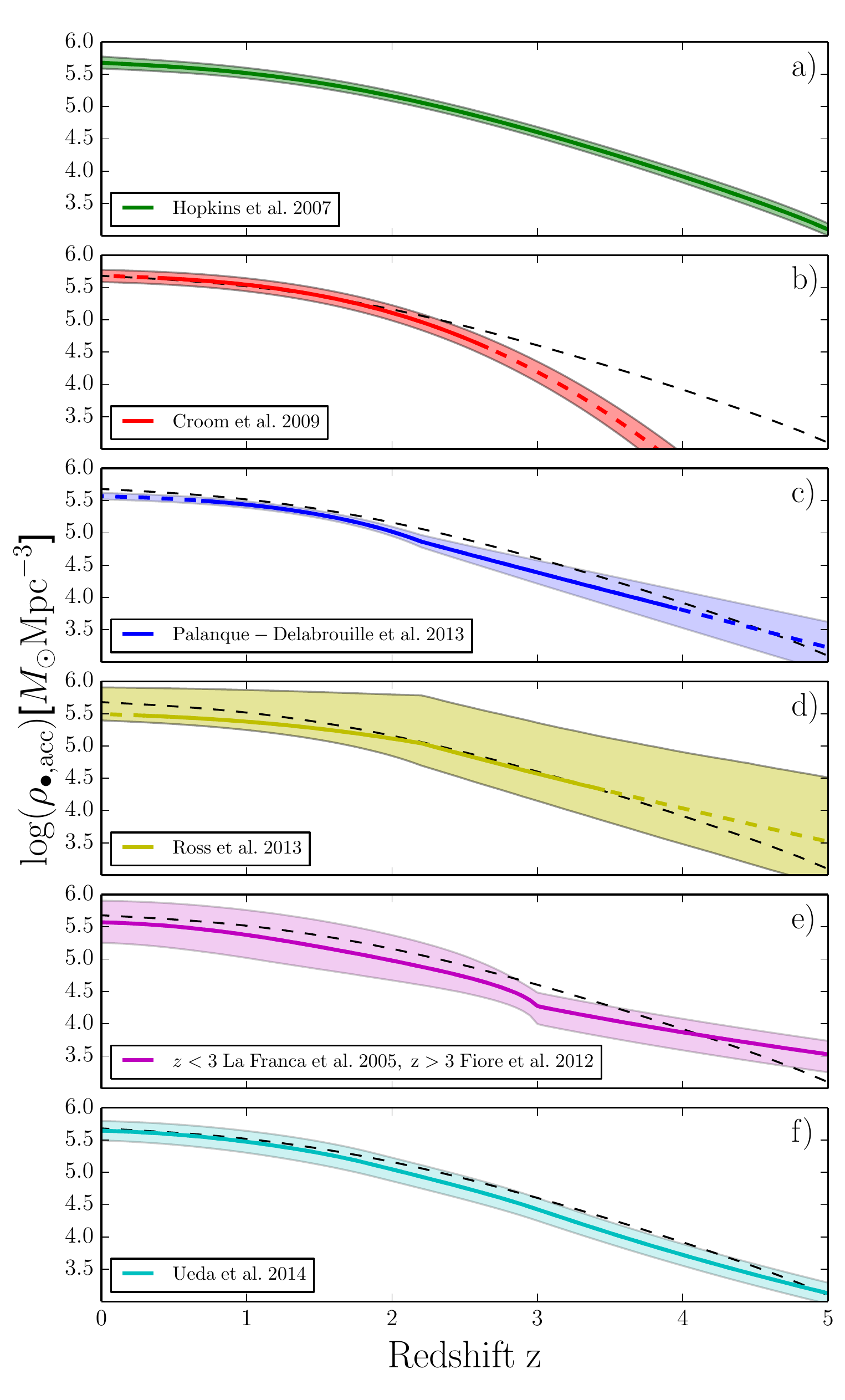}
\caption{We present the BHMD for the various QLFs adopted. The Monte Carlo median BHMD is shown as a solid colored line at redshifts where the QLF was estimated and as a dashed colored line at redshifts to which it was extrapolated. The colored areas correspond to the 68 percentile regions on the BHMD as a function of redshift. The dashed black line in panels b)-f) corresponds to the solid line in panel a) and serves as a reference for the eye.}
\label{figure_bhmd}
\end{figure}

Our results for the BHMDs are portrayed in Fig.\,\ref{figure_bhmd}\,a)-f). The solid colored lines show the Monte Carlo mean BHMD at the redshifts at which the QLF was determined. The dashed colored lines indicate the redshift regimes to which we extrapolated to. We only display the redshift evolution up to $z=5$, the maximum redshift for our SMD results. The light colored areas depict the 68 percentile regions estimated from the Monte Carlo method. 


\textbf{a)} The BHMD estimated from the bolometric QLF of \citet{Hopkins2007} (Fig.\,\ref{figure_bhmd}\,a) shows small uncertainties over the full redshift range. We compare our result for the local ($z=0$) BHMD with the result in the original publication in Table\,\ref{table_bhmd}. The results show excellent agreement. The small difference in the uncertainties might be a consequence of using different integration limits. 

We caution against over interpreting the high redshift evolution ($z>4$) of this BHMD, since only the bright end of the underlying QLFs was determined at these redshifts. This introduces large systematic uncertainties likely underestimating the BHMD at high redshifts.

\textbf{b)} The second BHMD is calculated using the QLF of \citet{Croom2009} (Fig.\,\ref{figure_bhmd}\,b).  
This BHMD falls steeply at high redshifts in contrast to all other evaluated BHMDs. It is most likely the effect of extrapolating the best fit QLF to redshifts beyond $z=2.6$, where it is not constrained by any data. We should exercise caution with these high redshift results.
On the other hand the local value of this BHMD fits well into the range of values estimated by other authors (Table\,\ref{table_bhmd}).

\textbf{c)} The results estimated from the optical QLF of \citet{Palanque2013} in Fig.\ref{figure_bhmd}\,c) show increasing uncertainties to higher redshift, while the local value has the smallest uncertainties of all five calculated BHMDs (see Table\,\ref{table_bhmd}). 

\textbf{d)} The fourth BHMD (Fig.\,\ref{figure_bhmd} d) calculated from the near-IR QLF of \citet{Ross2013} shows asymmetric uncertainties in the low redshift regime that increase towards higher redshifts. Extrapolation to large redshifts is needed since the QLF was determined using data in the redshift range $0.3 < z < 3.5$. Nonetheless the high redshift evolution agrees well with all other BHMDs except the one from \citet{Croom2009} (Fig.\,\ref{figure_bhmd} b). 
The $z=0$ BHMD is small compared to other values (Table\,\ref{table_bhmd}), possibly a result of having to assume the obscuration correction for the $B$-band instead of a near-IR correction.

\textbf{e)} The BHMD determined using the hard X-ray QLF of \citet{LaFranca2005} and \citet{Fiore2012} is not extrapolated for the redshift range shown. At redshift $z=3$ both QLFs are joined, which is evident from the change in the slope of the BHMD. This BHMD has high uncertainties at low redshifts, where the local BHMD value falls well in the range of determined BHMDs in this and other works. 

\textbf{f)} The sixth BHMD estimated from the QLF of \citet{Ueda2014} arrives at a local value close to the other estimated BHMDs above. It's redshift evolution is also similar, while the uncertainties show medium large values at all redshifts.

Our results for the local BHMD agree well with previous determinations from other authors (Table\,\ref{table_bhmd}). The overall BHMD evolution seems to be roughly consistent throughout all panels, but panel b) which is based on the adopted QLF of \citet{Croom2009}.

\begin{table}[h]
\centering
\caption{BHMD at $z=0$ Calculated from the Adopted QLFs (Upper Section) and Compared to Previous Investigations (Lower Section).}
\label{table_bhmd}
\begin{tabular}{ccc}
\tableline
\tableline \\
&$\log(\rho_{\bullet,\rm{acc}})(z=0)$ & QLF of\\
& $[M_\odot\rm{Mpc}^{-3}]$& \\
\tableline
a)&$5.68^{+0.10}_{-0.09}$ & \citet{Hopkins2007} \\
b)&$5.68^{+0.09}_{-0.10}$ & \citet{Croom2009} \\
c)&$5.57^{+0.05}_{-0.05}$ & \citet{Palanque2013} \\
d)&$5.49^{+0.42}_{-0.10}$ & \citet{Ross2013} \\
e)&$5.57^{+0.33}_{-0.32}$ & \citet{Fiore2012,LaFranca2005} \\
f)&$5.64^{+0.15}_{-0.15}$ & \citet{Ueda2014} \\
\tableline
& & Reference\\
\tableline
&$5.51$ & \citet{LaFranca2005}\\
&$5.68 \pm 0.10$ & \citet{Hopkins2007} \\
&$5.46^{+0.07}_{-0.08}$ & \citet{Yu2002} \\
&$5.66^{+0.15}_{-0.16}$ & \citet{Marconi2004} \\
\tableline
\end{tabular}
\end{table}

\section{The SMD-BHMD relation} \label{sec_relation}

\subsection{Correlation Between the BHMD and the SMD}

\begin{figure*}[h]
\centering
\includegraphics[width = 0.97\textwidth]{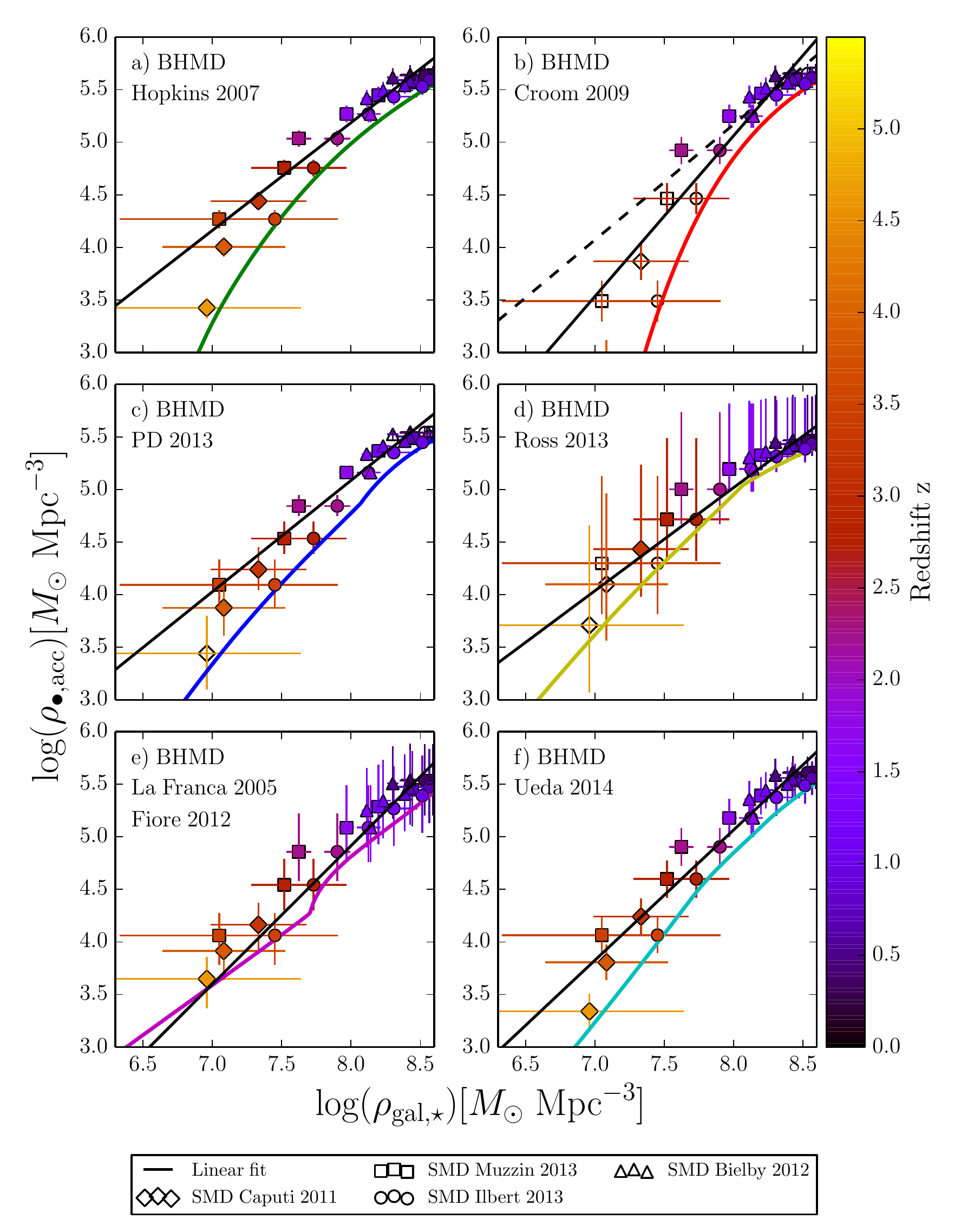}
\caption{The data points show the BHMD plotted against SMD, colored as a function of redshift. Each panel shows all SMD data versus one of the five BHMD estimations ( a) \citet{Hopkins2007}, b) \citet{Croom2009}, c) \citet{Palanque2013} (PD 2013), d) \citet{Ross2013}, e) \citet{Fiore2012,LaFranca2005}, f) \citet{Ueda2014}). The error bars show the associated 68 percentile regions/$1\sigma$ uncertainties. The colored solid lines show the SMD from the SFRD \citep{Madau2014} versus the specified BHMD. Their color code follows Fig.\,\ref{figure_bhmd} and has no relation to the redshift colored bar.  We display the best fit MCMC linear relation as the solid black line (Table\,\ref{table_smdbhmdrelation}). The dashed line in panel b) is the MCMC linear fit to the data with $z<3$.}
\label{figure_smdbhmdrelation}
\end{figure*}

In Fig.\,\ref{figure_smdbhmdrelation} a)-f) we show all values for the SMDs mapped onto one BHMD as a function of redshift. The data points are colored according to their redshift with their error bars resembling the uncertainties shown in Fig.\,\ref{figure_smd} and Fig.\,\ref{figure_bhmd}.

Up to intermediate redshifts, $z\approx 3$, we identify a linear relation of the data points in all double logarithmic plots, corresponding to a power law relation for the linear quantities. At low redshifts, $z<1$, the BHMD seems to saturate while the SMD still increases.
We further add the SMD values based on the SFRD of \citet{Madau2014} and map them onto the different BHMDs. The result is the solid colored line in each of the panels. According to the general offset to higher SMDs, the colored line lies beneath the data points by $\approx 0.4\,\rm{dex}$. Its shape roughly follows the data points in each panel, dominated by the different BHMD.

Assuming a simple power law ($\rho_{\bullet,\rm{acc}} = A \cdot \rho_{\star,{\rm{gal}}}^\beta$), without a redshift evolution for the normalization $A$, for the relation between SMD and BHMD, we fit the data points using the Markov Chain Monte Carlo (MCMC) algorithm of \citet{Kelly2007}. We include the uncertainties in both quantities as well as allow for scatter in the regression relationship. We assume that the intrinsic distribution of independent variables can be modeled as a mixture of three Gaussian functions. The posterior distributions are sampled using the Metropolis-Hastings algorithm option with at least ten thousand iterations. From the posterior distributions we calculate the median value as our best fit parameter and the 68 percentile regions as our $1\,\sigma$ uncertainties.
The best fitting curves using the median parameters for slope and intercept (see Table\,\ref{table_smdbhmdrelation}) are shown as the solid black lines in Fig.\,\ref{figure_smdbhmdrelation}.
We find that there is little to no intrinsic scatter in all fits, which is probably a consequence of the large errors on the data points.

Except in the case of Fig.\,\ref{figure_smdbhmdrelation}\,b) the fit is a good representation of the data. In this particular case the data points drop steeply towards lower SMDs due to the steep BHMD evolution at high redshifts of the \citet{Croom2009} QLF. We would like to caution against over interpreting this result at $z>3$ since the underlying QLF was only determined up to $z=2.6$, a lower redshift than any of the other adopted QLFs.

The best fitting values with uncertainties are displayed in Table\,\ref{table_smdbhmdrelation}. The uncertainties on the slope and the intercept are correlated with higher uncertainties in the slope resulting in much higher uncertainties in the intercept.

The relations including the BHMDs based on \citet{Hopkins2007,Palanque2013,Ross2013} show a slope close to unity and an intercept around $\sim -3.0$. The large uncertainties in the BHMD estimated from the QLF of \citet{Ross2013} give rise to large uncertainties in the fit parameters, compared to the other BHMDs.
The fit of the relation between the SMDs and the BHMD based on \citet{Croom2009} has the largest slope of $1.53$ and a corresponding small intercept of $-7.15$. If we restrict the fitting process to values with $z<3$, where the QLF was actually determined, we obtain a median slope of $1.10$ and intercept around $-3.61$. These agree with the fit parameters of the other data sets.

The relations with BHMDs calculated from \citet{LaFranca2005,Fiore2012,Ueda2014} show slopes around $1.2-1.3$ and corresponding smaller values for the intercept.

We have learned that the BHMDs and SMDs do follow a power law relation at least up to redshifts of $z\approx3$ where the error bars become too large to confidently constrain the relation. It is well represented by a power law with a slope around unity, with its individual value depending strongly on the data set.

\begin{table}[t]
\caption{This table presents a summary of our MCMC best fit parameters of the BHMD-SMD relation assuming a linear function for the logarithmic quantities. We present the median value of the posterior distribution and the 68 percentile regions around the median for each of the six BHMDs calculated and for two redshift restricted samples of the BHMD based on the \citep{Croom2009} QLF. Our results are contrasted by the local black hole mass to bulge mass relations of \citet{Kormendy2013} and \citet{Haering2004}.}
\label{table_smdbhmdrelation}
 \centering
 \def\arraystretch{1.4}
 \begin{tabular}{cccc}
 \tableline 
 \tableline
  & BHMD inferred from the QLF of & Intercept & Slope \\
  \tableline
 a)& \citet{Hopkins2007} & $-3.00_{-0.80}^{+0.92}$ & $1.02_{-0.11}^{+0.10}$ \\
 b)& \citet{Croom2009} & $-7.15_{-1.18}^{+1.53}$ & $1.53_{-0.19}^{+0.14}$ \\
 &\citet{Croom2009} $z< 3.0$ & $-3.61_{-1.16}^{+1.23}$ & $1.10_{-0.16}^{+0.13}$ \\
c) &\citet{Palanque2013} & $-3.38_{-0.89}^{+0.98}$ & $1.06_{-0.11}^{+0.12}$\\
 d)&\citet{Ross2013} & $-2.82_{-3.70}^{+3.36}$ & $0.98_{-0.41}^{+0.45}$ \\
  e) &\citet{Fiore2012},  &  & \\
  &\citet{LaFranca2005} & $-5.61_{-1.91}^{+2.04}$ & $1.32_{-0.25}^{+0.23}$\\ 
   f) &\citet{Ueda2014} & $-4.84_{-1.19}^{+1.43}$ & $1.24_{-0.18}^{0.14}$\\ 
  \tableline
  &\citet{Kormendy2013} & $-4.06\pm0.05$ & $1.16\pm0.08$ \\
  &\citet{Haering2004} &  $-4.12 \pm0.10$ & $1.12\pm0.06$ \\
  \tableline
 \end{tabular}
\end{table}

\subsection{The Cosmic Evolution of the Ratio of BHMD to SMD}

\begin{figure*}[h]
\centering
\includegraphics[width = 0.95\textwidth]{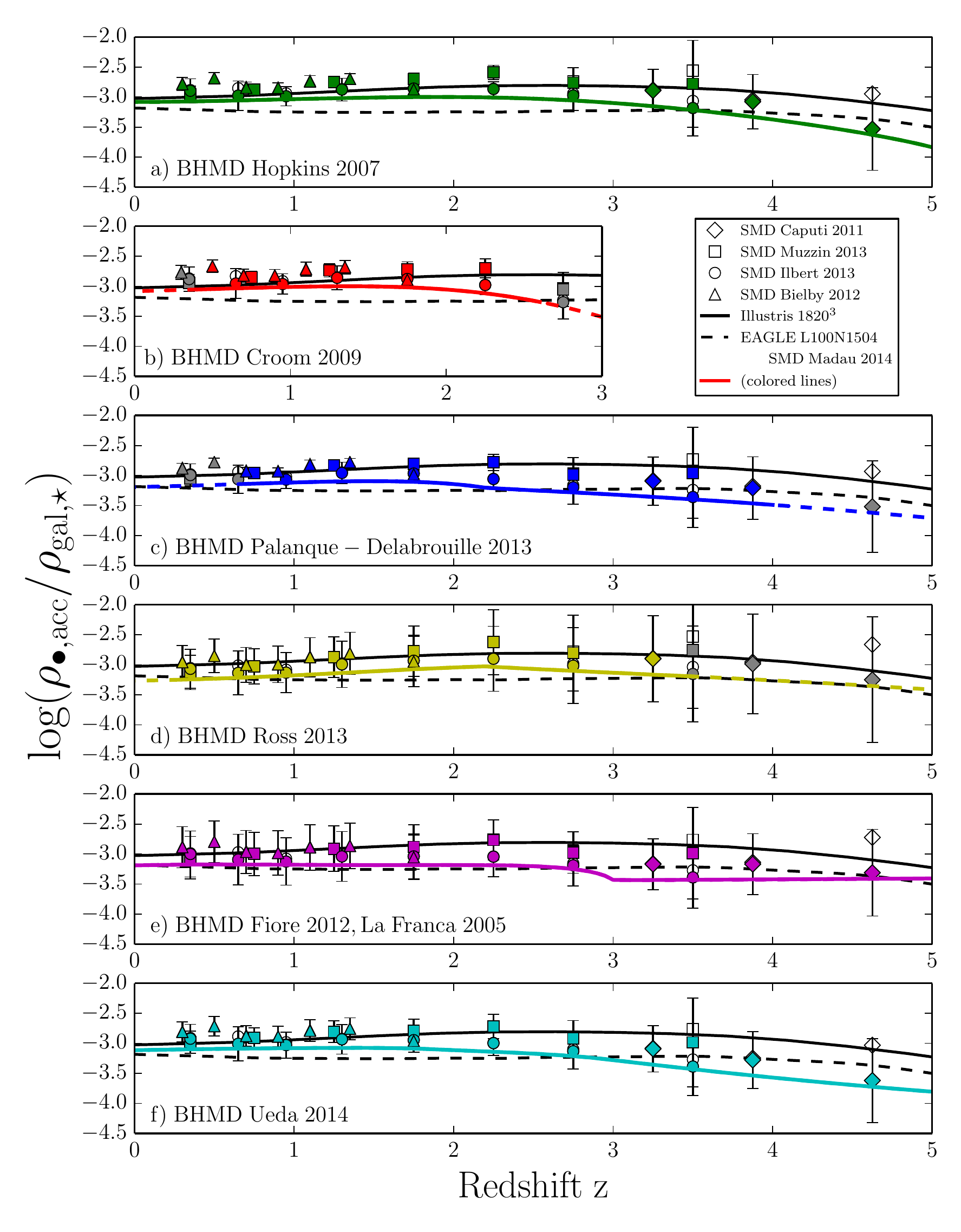}
\caption{The data displays the ratio of BHMD to SMD as a function of redshift. The data points refer to the SMD data calculated from SMFs (triangles \citep{Bielby2012}, squares \citep{Ilbert2013}, circles \citep{Muzzin2013}, diamonds \citep{Caputi2011}). The panels and colors refer to the different BHMDs as ordered in Fig.\,\ref{figure_smdbhmdrelation}. The lines (dashed/solid) represent the SMD from the instantaneous SFRD \citep{Madau2014}. Wherever we had to extrapolate the QLFs in redshift to calculate the BHMD the data points are grayed out and the line is dashed. The open data points refer to the average SMD values (see Table\,\ref{tab_SMDresults}), to point out the differences between them and the MC values, especially at high redshifts. The black solid line shows the BHMD to SMD ratio of the Illustris simulation with a resolution of $1820^3$ \citep{Sijacki2015}. The black dashed line shows the BHMD to SMD ratio of the EAGLE simulation (\citet{Furlong2015}, \citet{Rosas-Guevara2016}). }
\label{figure_bhmd_div_smd}
\end{figure*}

We now turn to analyze the redshift evolution of the logarithmic ratio of BHMD to SMD. 
Our results are summarized in Fig.\,\ref{figure_bhmd_div_smd}\,a)-f).
Every panel of the figure shows the ratio of both quantities as a function of redshift. 

If the BHMD and the SMD actually follow a simple power law relation with a slope of unity but with a redshift dependent normalization, this figure should show us any evolution of the normalization of this relation.

The colored data points correspond to the values of all SMDs in their respective redshift bins divided by one of the five different estimations of the BHMD. We show the average SMD (see Table\,\ref{tab_SMDresults}) as empty symbols while the MC mean SMD is over plotted in grey/color. We want to point out that they disagree at the highest redshifts which could possibly lead to a different interpretation.

The colored lines refer to the SMD calculated from the instantaneous SFRD \citep{Madau2014} divided by the BHMD, which underlines the influence of the BHMD across all panels.
Whenever the BHMD was estimated by extrapolating the underlying QLF we have grayed out the data points and show a dashed instead of a solid colored line.

We restrict panel b) to redshifts $z<3$ since the QLF of \citet{Croom2009} was only determined up to $z=2.6$.

According to the data points there is basically no redshift evolution of the logarithmic ratio within the $1\sigma$ error bars across all panels.

The BHMD to SMD ratio reaches it's maximum value at a redshift of $\sim2.2$ across all panels. Although the signature of the maximum in the generally flat evolution is very faint, it does coincide
with the maximum of merger activity and star formation.

Beyond redshifts of $z=3$ panel a),c)-f) show that the BHMD-SMD ratio slightly declines. Yet, given the large uncertainties at these redshifts the redshift evolution can still be considered consistent with being flat.

However, taking in account the error bars, our data does not support a rise of the ratio towards higher redshifts as might be expected if BH growth preceded galaxy stellar mass growth for the whole population.

\section{Discussion} \label{sec_discussion}

\subsection{Interpretation of the results}

In this work we use the mass densities of stellar matter and BHs to investigate co-evolution over a wide redshift range.

The SMD is the integrated average stellar mass per $\rm{Mpc}^3$ and as such does not provide any information about the underlying mass distribution. It traces foremost the galaxies which dominate the galaxy mass distribution, which are predominantly $M_\star$ ($L_\star$) galaxies. However these correspond to different kinds of systems at different redshifts.

The BHMD represents the average BH mass in a volume of one comoving $\rm{Mpc}^3$. We derived this quantity using the Soltan argument and therefore assume that the bulk BH mass growth occurs in quasar phases. Therefore the BHMD is naturally dominated by the most luminous AGN, which show the highest BH mass growth.

Although we have lost the information about the underlying BH and galaxy mass distributions, we have gained the ability to robustly estimate both quantities over a wide redshift range. 

The majority of the panels in Fig.\,\ref{figure_smdbhmdrelation} show a power law correlation between BHMD and SMD below $z=3$ with a slope around unity. The same signature can be seen in Fig.\,\ref{figure_bhmd_div_smd} where the logarithmic ratio of both quantities has remarkable little to no redshift evolution up to $z=3$. We interpret this as lock-step growth of the two quantities from $z=3$ to $z=0$.

While this result does not allow us to conclude that all individual BHs and their hosts grow in lock-step in this redshift range, it does strongly suggest that as a population, the systems, which dominate the mass growth, do follow this lock-step growth.

At higher redshifts around $3<z<5$ and given the completeness of the SMFs and QLFs, the BHMD to SMD ratio does not show a strong evolution. However, at the highest redshifts the ratio of BHMD to SMD seems to decrease slightly rather than to increase. While an increase could be interpreted as generally over massive BHs compared to the local value, a decrease would imply the opposite.

The evolution of the BH galaxy scaling relations has been a vivid topic of investigation.

While at redshifts below $z\sim1$ the BH to galaxy bulge mass relation does no seem to evolve \citep[e.g.][]{Cisternas2011b}, studies at higher redshift \citep[e.g.][]{Peng2006,Treu2007,Jahnke2009,Bennert2011} find a positively evolving relation, suggesting that larger BHs ($M_{\bullet} > 10^8\,M_\odot$) pre-date the formation of their host galaxy bulges at these redshifts. 
Most similar to our method, \citet{Zhang2012} estimate the evolution on the $M_\bullet-M_{\star,\rm{bulge}}$ relation using BHMDs and SMDs. Their results also favor a positively evolving BH to galaxy bulge mass relation.

However, simulations of merging disc galaxies \citep{Robertson2006, Johansson2009} seem to disfavor the picture in which SMBHs develop significantly before their parent bulges. 

Regarding the relation between the BH and the total galaxy stellar mass, \citet{Jahnke2009} find the BH mass to \textit{total} stellar mass relation not to be evolving, whereas \citet{Decarli2010} and \citet{Bennert2011} find a positive evolution.

\citet{Kormendy2013} illustrate in their Figure\,38 that on average $M_\bullet/M_{\star,\rm{bulge}}$ increases from its local value \citep{Haering2004} towards higher redshift. However, these high redshift data points suffer selection effects that are hard to quantify and sample bias towards more luminous systems. Also bulge disc decomposition becomes increasingly difficult in non-local systems with an AGN. If the total instead of the bulge stellar mass is used, the evolution of the ratio with z weakens or even disappears below $z=2$ \citep[see][Fig. 38b, green points]{Kormendy2013}.

\citet{Schulze2014} showed very convincingly that evolutionary trends in the apparent ratio of the BH to galaxy bulge mass relations do not carry over to an intrinsic offset once selection effects \citep{Lauer2007} are taken in account. They do not demonstrate this for the studies of \citet{Peng2006} and \citet{Decarli2010} above but for a variety of other works \citep[e.g.][]{Wang2010}).

In conclusion the BH to bulge mass relation seems to be evolving positively with redshift (see also \citet{Volonteri2016}), while the studies on BH to total stellar mass ratio are more ambiguous. 

While our work somewhat disagrees with \citet{Decarli2010,Bennert2011}, it fits well into the line of recent works \citep[see][]{Jahnke2009,Kormendy2013} finding only very little evidence for an evolution of the ratio of BH to \textit{total} galaxy stellar mass up to $z=3$.

On this basis we find that BH mass growth \textit{did not} precede \textit{total} stellar mass growth in any of our data sets. The data of systems that show over massive BHs compared to their galaxy properties could well be only a subset of the total distribution of BH and host galaxy systems.

\subsection{Comparison to hydrodynamical cosmological simulations}

We compare our results, which are based on observational data, with the results of the Illustris \citep{Vogelsberger2014, Genel2014} and the EAGLE \citep{Schaye2015} simulations.

The Illustris simulation is a large scale cosmological hydrodynamical simulation. It covers a volume of $106.5\,\rm{Mpc}^3$ with more than 12 billion resolution elements. State-of-the-art physical models describing the relevant physics for galaxy formation are included, allowing for the study of the properties black holes and their host galaxies.

For our purpose we use the BH accretion rate (BHAR) and star formation rate density (SFRD) of the highest resolution Illustris simulation ($1820^3$), published in \citet{Sijacki2015}.
Both quantities are integrated across redshift to calculate the BHMD and the SMD. We rescale the SFRD from a Chabrier to a Salpeter IMF and use a return fraction of 0.27, according to Salpeter IMF, to calculate the SMD.

The simulation calibrates the efficiency of the BH feedback to reproduce the observed cosmic star formation history and the $z=0$ stellar mass function. Furthermore, the quasar efficiency factor has been set to match the normalization of the $M_{\rm{BH}}-\sigma$ relation in isolated galaxy mergers. While the Illustris simulation slightly over predicts the absolute values of the total stellar and black hole mass density, their evolution follows the data calculated in this work and the SMD from the SFRD of \citet{Madau2014} closely. Regarding the SMD this is not surprising, since the BH feedback was tuned to reproduce the star formation history. The evolution of the BH mass density, however, is not pre-determined.

The EAGLE simulations \citep{Schaye2015, Crain2015} are a suite of hydrodynamical cosmological simulations run with a version of the GADGET 3 code that includes a new hydrodynamics solver, new time stepping, and new sub-grid physics. We adopt the SMD of \citet{Furlong2015} and the BHMD (Rosas-Guevara et al., in preparation) from the large simulation Ref-L100N1504 with $1504^3$ particles in a box of $100$ comoving $\,\rm{Mpc}^3$. While the Illustris simulation is designed to match the observed star formation history and therefore the redshift evolution of the feedback efficiency is calibrated, the EAGLE simulation only matches to $z\sim0$ observational data and thus both BHMD and SMD evolution are not pre-determined. The sub-grid physics concerning stellar and AGN feedback are calibrated to match the $z\sim0$, local galaxy stellar mass function as well as the normalization of the BH galaxy mass scaling relation, which is mainly determined by the efficiency parameter of AGN feedback. 

The SMD includes only stellar mass in galaxies excluding the mass associated with intra cluster light. The EAGLE simulation uses a Chabrier IMF and thus we have rescaled the mass density to a Salpeter IMF. The BHMD only includes the accreted black hole mass and not the original seed masses and is thus the best comparison to the BHMD we calculated.

The simulations and our data are computed assuming slightly different cosmological parameters. Yet, the maximum difference in terms of cosmological volumes at a redshift of $z=3$ are below $7\%$ and thus we are confident to compare them to another without recalibrating to one cosmological parameter set.

We compare the simulation data to our inferred BHMD to total SMD ratio in Fig.\,\ref{figure_bhmd_div_smd}. The data from the Illustris simulation is portrayed as a black solid line, while the data from the EAGLE simulation is shown as a black dashed line.

The Illustris BHMD to SMD ratio (black solid line) follows our data points closely and is almost always within the $1\sigma$ error bars of the data points.

The data from the EAGLE simulation, the black dashed line in Fig.\,\ref{figure_bhmd_div_smd}, matches our data points as well in panels c), d), e) and f) and at $z\gtrsim3$ in all panels.  

Both simulations agree well with our data and show a nearly flat redshift evolution of the BHMD to SMD ratio up to redshifts of $z>4$ where it starts to decline slightly.

\section{Conclusion} \label{sec_conclusion}

In this work we have calculated the BHMD and SMD to evaluate possible signatures from BH-galaxy co-evolution up to high redshifts. 
We calculated the SMD from multiple measurements of SMFs and the cosmic SFRD of \citet{Madau2014}, while we have used the Soltan argument on various QLFs to obtain measurements of the BHMD. The values of our calculated SMDs at various redshifts agree well with the values in the recent review of \citet{Madau2014}. We also compare our BHMD at $z=0$ with previous studies and find generally good agreement.

Our data shows clear lock-step growth of the BHMD with the SMD in the redshift regime ${z=0-3}$. If the BHMD-SMD-relations, shown in Fig.\,\ref{figure_smdbhmdrelation}, are fitted with power laws, we find values for the exponent between $1.0$ to $1.3$. These values are surprisingly close to unity. While local estimates of the BH mass to stellar bulge mass relation have very similar slopes, the SMD-BHMD cannot be easily compared to the local relations. It is an evolutionary sequence of integrated distributions of BH mass and stellar mass and not a ``snapshot'' of their distributions at $z=0$.

We show that the logarithmic ratio of BHMD to SMD seems to be roughly constant from $z=0$ to $z=5$ within the $1\sigma$ uncertainties. Therefore BH mass growth and stellar mass growth trace very well over long periods of time up to high redshifts. 

Since the considered quantities hold information about the integrated distribution of BHs and galaxies, information about the distributions itself cannot be accessed. Therefore we cannot distinguish between physical processes that connect or regulate BH and stellar mass growth and scenarios of secular evolution for both quantities. 

However, our data does not support a scenario in which BH growth preceded \textit{total} galaxy stellar mass growth. In light of this result, BHs, which are over massive according to BH galaxy scaling relations, are likely to be outliers of the full distribution.

From the results of the EAGLE and Illustris simulations, including galaxy and AGN feedback, we also do not expect the BHMD to SMD ratio to strongly evolve with redshift. 

At high redshifts measurements of BH mass and galaxy properties become increasingly difficult. On this basis following the local BH mass galaxy relations up to high redshifts to understand their evolution and origins is an involved task. With this work we try to open up other angles to investigate BH and galaxy co-evolution based on large data sets. 

Our effort is not free of assumptions and the underlying SMFs and QLFs do suffer completeness issues at the highest redshifts and the lowest masses/luminosities. But so far, using the same analysis technique on all data sets, our results are consistent across all used SMFs and QLFs.

Upcoming large scale surveys of galaxies and quasars will extend our knowledge of the SMFs and QLFs to even higher redshifts and allow us to trace back the possible co-evolution of BH and their hosts until the cosmic dawn.

\acknowledgements{We thank the anonymous referee for a constructive review. Further we would like to thank Knud Jahnke, Gurtina Besla and Dan Stark for in depth, helpful discussions and suggestions to this work. We furthermore would like to thank Debora Sijacki, Yetli M. Rosas-Guevara, Michelle Furlong and Joop Schaye for providing the Illustris and EAGLE simulation data. JTS and XF thank supports from the NSF grants AST 11-07682 and 15-15115.}

\begin{table*}[t]
\caption{Summary of the Stellar mass functions adopted in this work.}
\label{tab_smfsummary}
\centering
 \def\arraystretch{1.4}
\begin{tabular}{p{3cm} p{3.5cm} p{2cm} p{2cm} p{1.2cm} p{2cm} p{2cm}}
\tableline
Reference & Best fit &  z Range & Survey/Field & Area & $\rm{N}_{\rm{Galaxies}}$ & IMF  \\
\tableline
 \cite{Caputi2011} & their Table\,2 &  $z = 3-5 $ & UKIDSS \& UDS & $0.6\,\rm{deg}^2$ & $1213$ & Salpeter \\
 \cite{Ilbert2013} & their Table\,2 full sample & $z = 0.2-4.0 $ & UltraVista DR1 / Cosmos & $1.52\,\rm{deg}^2$ & $220000$ & Chabrier \\  
 \cite{Muzzin2013} & their Table\,1, sample: all, first redshift bin: third row, all other bins: first row, in reference &  $z = 0.2-4.0 $ & UltraVista DR1 / Cosmos  & $1.62\,\rm{deg}^2$ & $95675$ & Kroupa  \\ 
 \cite{Bielby2012} & their Table\,7 & $z = 0.2-2.0 $ & WIRCam Deep Survey & $2.03\,\rm{deg}^2$ & $\sim 50000$ per band & Chabrier\\ 
\tableline
  \multicolumn{7}{c}{ 
\begin{minipage}{2.\columnwidth}
\footnotesize \textbf{Notes.} All references use a $\Lambda$CDM cosmology with $h = 70, \Omega_{\rm{m}} = 0.3,  \Omega_{\Lambda} = 0.7$.  
\end{minipage}}
\end{tabular}
\end{table*}

\begin{table*}[t]
\scriptsize
\caption{Summary of the Quasar luminosity functions adopted in this work.}
\label{tab_qlfsummary}
\begin{tabular}{p{3cm} p{2cm} p{3cm} p{5.5cm} p{1.5cm} p{1.2cm}}
\tableline
Reference & z Range & Survey/Field & Luminosity Range & Area & $\rm{N}_{\rm{AGN}}$ \\
\tableline
 \cite{Hopkins2007} & $z = 0-6$ & compilation, see Table\,1 in \cite{Hopkins2007} & $ 41 \leq \log(L_{\rm{bol}}/[\rm{erg}\,\rm{s}^{-1}]) \leq 49 $ &  & $\sim 49000$ \\ 
 \cite{Croom2009} &  $z = 0.4 -2.6$ & 2dF-SDSS LRG, 2SLAQ & $M_{\rm{g}} (z=2) < -21.5$ & $191.9\,\rm{deg}^2 $& 10637 \\  
 \cite{Ross2013}&  $z = 0.3 - 3.5$ & SDSS-III: BOSS  & $M_{\rm{i}} (z=2.2) < -24.5$ & $2236\,\rm{deg}^2 $& $23201$  \\ 
 \cite{Palanque2013} & $z = 0.68 - 4.0$ & SDSS-III: BOSS Str. 82 + MMT & $ M_{\rm{g}}(z=2) < -22.3$ &  $14.5\,\rm{deg}^2 $ & $1877$ \\ 
 \cite{Fiore2012} & $z = 3-7$ & 4Ms CDFS & $42.75 < \log(L(2-10\rm{keV})/[\rm{erg}\,\rm{s}^{-1}]) < 44.5$ & $464.5\,'^2$ & $54$  \\ 
 \cite{LaFranca2005} & $z = 0-4$ & HELLAS2XMM, also see Table\,1 in \cite{LaFranca2005}& $42 \lesssim \log(L(2-10\rm{keV})/[\rm{erg}\,\rm{s}^{-1}]) \lesssim 46$ & $3\,\rm{deg}^2$ + others & $508$ \\ 
 \cite{Ueda2014} & $z = 0-5$ & compilation, see Table\,1 in \cite{Ueda2014}& $42 \lesssim \log(L(2-10\rm{keV})/[\rm{erg}\,\rm{s}^{-1}]) \lesssim 46$ & see Table\,1 in \cite{Ueda2014}& $4039$\\
 \tableline
 \multicolumn{6}{c}{ 
\begin{minipage}{2.\columnwidth}
\footnotesize \textbf{Notes.} All references use a $\Lambda$CDM cosmology with $h = 70, \Omega_{\rm{m}} = 0.3,  \Omega_{\Lambda} = 0.7$ except for \cite{Palanque2013} which use slightly different cosmological parameters ($h = 71,  \Omega_{\rm{m}} = 0.267, \Omega_{\Lambda} = 0.734$).   
\end{minipage}}
\end{tabular}
\end{table*}

\bibliographystyle{apj}

\end{document}